\documentclass[prb,twocolumn,aps,showpacs]{revtex4}
\usepackage{graphicx}%
\usepackage{dcolumn}
\usepackage{amsmath}
\usepackage{amssymb}
\usepackage{epsfig}

\begin{document}

\title{Cavity polaritons in the presence of symmetry-breaking disorder: closed-path time formalism }

\author{Z. Koinov}
\affiliation{Department of Physics and Astronomy, University of
Texas at San Antonio,
San Antonio, TX 78249,USA}%
\email{Zlatko.Koinov@utsa.edu}

\begin{abstract}

According to the mean-field theory of Zittartz, when subject to a
symmetry-breaking disorder, the order parameter and the energy gap
of an excitonic insulator are gradually suppressed up to a
critical disorder strength. Recently, Marchetti, Simons, and
Littlewood have used a replica trick to investigate the effects of
disorder on the condensation of cavity polaritons. Within their
nonlinear sigma model, it was found that  the saddle-point
equations assume the form reported previously by Zittartz in the
contest of the symmetry broken excitonic insulator, but with an
order parameter, to which both photons and excitons contribute. In
this paper, we apply a closed-path time Green's function approach
as an alternative to the replica technique to formulate a
nonperturbative description of cavity polaritons in the presence
of a symmetry-breaking disorder. A field theoretical method is
used to derive the Schwinger-Dyson equations for the average
photon field and the average single-particle Green's function. In
contrast with the nonlinear sigma model and the corresponding
saddle-point equations, we obtain that the exact Schwinger-Dyson
equations cannot be mapped to the corresponding equations derived
by Zittartz. This result not only shows that the theory of
Zittartz  cannot be applied to the excitons in a disordered
quantum well coupled to the cavity photons with only minor
modifications, but arises a question about the validity of the
replica trick as well.
\end{abstract}

\maketitle
\section{\label{sec:sec1}Introduction}
The phenomenon of the Bose-Einstein condensation (BEC) in atomic
gases and superconductors attracts much attention in recent years.
Substantial efforts has been recently devoted to BEC of excitons in
quantum wells (QW) and microcavities (MC) \cite{But}. QW embedded
within semiconductor MC have attracted considerable interest
\cite{S,LW} due to the following two reasons. Firstly, the recent
progress in the growth and manipulation techniques of semiconductor
heterostructures allows us to control the coupling between photons
and excitons. The QW excitons embedded in semiconductor MC may be
found in either weak- or strong-coupling regimes. In what follows we
assume the strong coupling regime, where the photon-exciton
interaction is larger than the exciton and photon damping rates, and
therefore, the normal modes are mixed exciton-photon modes, called
cavity polaritons. Secondly, it is expected that the cavity
polaritons should have bosonic
behavior, and so are candidates for Bose condensation.\\
 Despite the progress made in semiconductor technology, a weak
disorder may exist due to the following reasons: the interface
roughness, the local thickness fluctuations during crystal-growth
processes, randomly distributed impurities, boundary
irregularities, and fluctuations of the alloy concentration of the
epitaxial layers. The disorder causes additional problems in the
theory because all quantities of interest depend on the
corresponding random potential, created by the disorder. The
actual potential is unknown, but it is not important for the
physical properties. Instead, the disorder is considered by means
of the probability distribution of the random potential, i.e. one
should perform the averaging over all possible random
potentials.\\
Turning our attention to the theoretical situation, we find that
some authors\cite{LW} have focused on the model, which assumes
that the excitons are localized by the disorder, and can be
described as two-level oscillators coupled to the light.  In this
model the original electron-hole-photon Hamiltonian is reduced to
the Hamiltonian, which describes the so-called generalized Dicke
model.\cite{D} The model is valid only for describing the very low
energy excitonic states, and cannot be applied to a
symmetry-breaking disorder, because it assumes that there exists
only a symmetry preserving disorder
potential, which is strong enough to localize the excitons. \\
More complicated approach to the problem of excitons (or excitonic
polaritons) in weakly disordered semiconductors is based on the
assumption that the disorder affects only the center-of-mass
motion, but does not affect the exciton internal degrees of
freedom (see, e.g., Ref. [\onlinecite{R}] and references therein).
Recently, this idea  has been applied to the cavity
polaritons.\cite{M} According to this
disorder-affected-center-of-mass-motion (DACMM) approach, the
two-particle Schrodinger equation for an isolated exciton
separates into two equations: the Wannier equation for the
relative motion of the electron-hole pair and the Schr\"{o}dinger
equation for the exciton center-of-mass motion in a random
potential. As a result, the two-particle exciton wave function can
be factorized, and therefore, the coupling strength of an exciton
to light is a random quantity, which depends on the exciton
center-of-mass eigenfunctions $\Psi_i(\bf{R})$. The next step in
the DACMM approach is to use numerical simulations to generate
random potentials. Once the energies $E_i$ and eigenfunctions
$\Psi_i(\bf{R})$ are calculated for a particular random potential,
the radiative decay rates, the absorption (the optical density),
or the exciton-photon coupling strengths can easily be evaluated
numerically on a grid of a given number of points. There exists a
many-body version of DACMM approach,\cite{B} where instead of
numerical simulations, the Green's function of the center of mass
of an isolated exciton in the random field is calculated in the
coherent potential approximation. It is expected that the
factorization of the wave function is justified  in the very low
density regime. Strictly speaking, the DACMM approach is based on
the assumption that the disorder and the interactions (Coulomb and
electron-photon interactions) could be treated independently. In
other words, the factorization assumption greatly simplifies the
problem, but it separates the disorder and the interactions, and
therefore, we may expect that the DACMM approach underestimates
the influence of the disorder and
may lead to incorrect conclusions.\\
Decades ago, Zittartz\cite{Zi} demonstrated that the disorder and
the interactions can be treated simultaneously if we perform the
averaging over the disorder in the beginning of all calculations.
Within this approach, the Green's functions are defined as
$\overline{<\widehat{T}\{....\}>}$ and the brackets $<...>$ denote
a thermal average, while $\overline{f}$ means the average of $f$
over the random potential created by the disorder. Zittartz
applied the Abrikosov and Gor'kov theory,\cite{AG} developed for
the case of superconductors in the presence of a symmetry-breaking
disordered potential, to the case of an excitonic insulator in the
presence of normal impurities. The theory of Zittartz can be used
with only minor modifications to investigate the effects of a
symmetry-breaking disorder potential on the two-dimensional
excitonic condensate in a high density regime, where the screened
Coulomb interaction could be replaced by a contact interaction
with a coupling strength $g_c$. Assuming a Gaussian disorder
potential with zero mean, and variance
$\overline{V(\textbf{r})V(\textbf{r}')}=\Lambda \delta
(\textbf{r}-\textbf{r}')$, one can obtain the following set of
equations for the order parameter $\Delta$ at the Fermi surface:
\begin{equation}\begin{split}
&\Delta= \frac{g_c}{\beta}\sum_{m=-\infty}^\infty
\frac{1}{\sqrt{1+u_m^2}},\\
&\frac{\omega_m}{\Delta}=u_m\left[1-\frac{\alpha
}{\sqrt{1+u_m^2}}\right].\label{Zeqs}\end{split}\end{equation}
Here $\alpha=2\Lambda m_{exc}/\Delta$ , $m_{exc}$ is the exciton
reduced mass, $\omega_m=(2m+1)\pi/\beta$, $\beta=(kT)^{-1}$, where
$T$ and $k$ are the temperature and the Boltzmann constant,
respectively. The Matsubara summation in (\ref{Zeqs}) must be
cutoff at the energy $\epsilon_0$, which depends on the chemical
potential $\mu$. The solution of the above equations shows that:
(i) the order parameter and the energy gap are gradually
suppressed up to a critical disorder strength; (ii) the
suppression of the energy gap is more rapid than that of the order
parameter, which means that the existence of a gapless condensed
phase is
possible. \\
Generally speaking, the case of cavity polaritons in the presence
of a symmetry-breaking disorder is more complicated than the
excitonic condensate, because the theory has to take into account
the photonic contributions to all quantities of interest. To the
best of our knowledge, there exists only one paper by Marchetti,
Simons, and Littlewood\cite{MSL} (MSL), where a model for cavity
polaritons in the high density regime in the presence of a
symmetry-breaking disordered potential is proposed, treating the
disorder and the interactions simultaneously. MSL have used the
so-called replica trick to perform the averaging over the
disorder. The replica trick\cite{EA} is based on the following
relationship: $\overline{\ln Z}=\lim_{N\rightarrow
0}\left[(\overline{Z^N}-1) / N\right]$, where $Z$ is the
generating functional. Once replicated, MSL have decoupled the
arising quartic term in $\overline{Z^N}$ by means of the
Hubbard-Stratonovich transformation with the introduction of a
matrix field $Q(\textbf{r},\imath \omega_m)$, and then,
integrating over the fermionic fields the problem is reduced to
the so-called nonlinear sigma-model action, previously used to
study superconductors with magnetic impurities.\cite{LS} The final
step in this approach is to draw conclusions by investigating the
structure of the saddle-point solution. At the level of the
saddle-point approximation the following three statements take place:\\
(i) while the chemical potential does not exceed the cavity edge
mode $\omega_c$, an order parameter $|\Delta|\neq 0$ is developed:
 \begin{equation}|\Delta|=g|\psi|+|\Sigma|.\label{MSL1}\end{equation}
Here, $g$ is the exciton-photon coupling constant, $\psi$ is the
average photonic field, $g|\psi|$ and $|\Sigma|$ are the photonic
 and the excitonic contributions to the order
parameter, respectively;\\ (ii) the excitonic order parameter
$|\Sigma|$ and the photonic field $|\psi|$ are not independent
quantities because of the following constrain:
  \begin{equation}(\omega_c-\mu)|\psi|=(g/g_c)|\Sigma|;\label{MSL2}\end{equation}
 (iii) the
saddle-point equations in the high density regime can be mapped to
the corresponding set of equations by Zittartz (\ref{Zeqs}), but
with the order parameter $|\Delta|$, defined by (\ref{MSL1}), and
with $g_c$ replaced by $g_{eff}=g_c+g^2/(\omega_c-\mu)$. Because
of the correspondence between the Zittartz's equations and the
saddle-point equations, MSL have concluded that in the low-density
regime, where the excitations are mainly excitonic like (only a
small fraction of photons contributes to the condensate), the
order parameter and the energy gap are gradually suppressed up to
a critical strength of the disorder. The suppression of the energy
gap is more rapid than that of the order parameter, and therefore,
the existence of a gapless condensate is possible. When the
density of the excitations is increased, the chemical potential
rises (first linearly with the density), and when the chemical
potential approaches $\omega_c$, the excitations become photonic
like. In other words, the character of the condensate changes from
being excitonic to photonic.
\\
The purpose of this paper is to show that all of the above
conclusions are drawn only because MSL have performed the
averaging over the disorder using the replica trick. To justify
our point, we shall treat the effects of a symmetry-breaking
disorder on the cavity polaritons by applying the closed-path time
(CPT) (or Keldysh) Green's function technique\cite{K}. The main
reason for using this approach is that in the case of a static
random potential the Keldysh closed contour in the time direction
leads to an automatically disorder independent generating
functional. In other words, the CPT approach allows us not only to
avoid the need to introduce replicas, but to perform the averaging
over the disorder in the beginning of all calculations as well,
which is the main requirement when the disorder and the
interactions are treated simultaneously. The special form of the
Keldysh time contour automatically ensures that the denominator in
the representation of the Green functions via functional integrals
is equal to unity. The last allows us to derive the exact
equations for the average photon field and for the average
single-electron Green's function. In the quantum-field theory,
these equations are known as the Schwinger-Dyson (SD) equations.
In contrast with the saddle-point equation (\ref{MSL1}), the exact
SD equations clearly indicate that in the presence of a
symmetry-breaking disorder the photonic contribution to the mass
operator is not proportional to the average photon field. We shall
see that the results derived by applying the replica trick
correspond to the assumption that one can replace the average of
the product of two random functions with the product of the
corresponding average functions. However, it is known that the
average of the product is not the product of the averages, and
therefore, the validity of the conclusions based on the replica
trick is questionable. The exact result that the photonic
contribution to the mass operator is not proportional to the
average photon field does not allow us: (i) to map the SD
equations to the corresponding Zittartz equations; (ii) to use the
Ward identities in order to prove the existence of the Goldstone
mode below the critical temperature, and therefore, the question
about the existence of a condensate of cavity polaritons in the
presence of a symmetry-breaking disorder
remains open. \\
The remainder of the paper is organized as follows. In Sec. II, we
discuss the formation of a condensate in the absence of a
disorder. This is because we intend to check the validity of the
saddle-point approximation by eliminating the effects generated by
the  replica trick. By applying the Matsubara Green's function
method we demonstrate that the polariton spectra can be obtained
from the common poles of the photon and the two-particle
electron-hole Green's functions. It turns out that the
saddle-point equations in the absence of a disorder lead to the
same conclusions as those drawn by applying the Matsubara Green's
function method. The approach used in Sec. II is very general and,
in principle, it is able to treat any density regimes. It also
allows us to prove the existence of the Goldstone mode below the
critical temperature. Our method provides a set of coupled BCS and
Bethe-Salpeter (BS) equations similar to the corresponding
equations for an excitonic condensation.\cite{G,Z} Due to the
photonic contribution to the condensate, the BCS and the BS
equations are more complicated than the equations reported in our
previous paper\cite{Z}, and it would be a very challenging task to
solve them in the case of a low-density limit. Such an ambitious
task will be left as a subject of future research. In Sec. III we
treat the effects of a symmetry-breaking disorder on the cavity
polaritons
  by applying CPT Green's function
technique, because this approach is analytical nonpertubative one
which provides exact results. Furthermore, the Keldysh formalism
could be applied to the cavity polaritons in nonequilibrium
conditions. \\

\section{\label{sec:sec2}Cavity polaritons in the absence of a disorder  - the Matsubara Green's function approach}
The system under consideration consists of a single QW grown
inside a semiconductor MC is an arrangement of two-plane parallel
mirrors with reflectivity close to unity. The two infinite and
parallel perfect mirrors are perpendicular to $z$-axis, separated
by a distance $L_0$, one mirror is at $z=L_0/2$, and the other at
$z=-L_0/2$. In what follows we are interested in the case of a
single QW extending over $-L/2<z<L/2$ made from a direct-gap
semiconductor with nondegenerate and isotropic bands when the
electron-hole motion along the z-direction is confined between two
parallel, infinitely high potential barriers.  With the perfect
confinement approximation the dispersion laws for electrons and
holes are ${E}_c(\textbf{k}_c,\lambda)=\textit{E}_g
+\textbf{k}_c^2/2m_c+\pi^2\lambda^2/2m_cL^2$ and
${E}_v(\textbf{k}_v,\xi)=-\textbf{k}_v^2/2m_v-\pi^2\xi^2/2m_vL^2$,
respectively. Here $m_c$ ($m_v$) is the electron (hole) effective
mass, $E_g$ is the energy gap, and $\textbf{k}_{c,v}$ is a
two-dimensional (2D) wave vector. $ \lambda,\xi = 1,2, . . .$
denote the quantum number of the states in the infinitely deep
wells. In what follows we use the simplest approximation which
takes into account only the first electron and hole confined
levels, i.e. $\lambda=\xi=1$.\\ For each photon wave vector there
are two possible polarizations: one with transverse electric field
(TE), and second, with transverse magnetic field (TM). \cite{ZR}
In what follows we will take into account only the TE modes which
interact with transverse polarized excitons. The longitudinal
photon modes mediate the Coulomb interaction between the charges
in the QW, but we neglect this effect assuming that the Coulomb
interaction between the charges in the QW is affected only by the
confinement of the charges. The cavity-mode dispersion is
$\Omega_s(\textbf{q})=c\sqrt{q^2+(\pi s/L_0)^2}$, where $s=1,
2,...$, and $\textbf{q}$ is a 2D vector. In the following, we
suppose that the only $s=1$ cavity modes $\Omega(\textbf{q})$
interact with the electron system. \\In terms of the field theory,
the transverse and the longitudinal photon modes are described by
boson fields $A_\perp(\rho)$ and $A_\parallel(\rho)$,
respectively. They interact with the electron system, described by
fermion fields $\psi^+(y)$ and $\psi(x)$. The total action of the
system is $$S= S^{(e)}_0+S^{(\omega)}_0+S^{(e-\omega)}.$$ The
actions for non-interacting electrons and photons are
$$S^{(e)}_0=\overline{\psi} (y)G^{(0)-1}(y,x)\psi (x),$$ and
\begin{equation}\begin{split}&S^{(\omega)}_0=\frac{1}{2}A_{\parallel}(\rho)D^{(0)-1}_{\parallel}(\rho,\rho')A_{\parallel}(\rho')
\\&+\frac{1}{2}A_{\perp}(\rho)D^{(0)-1}_{\perp}(\rho,\rho')A_{\perp}(\rho'),\nonumber\end{split}\end{equation}
respectively. The electron-photon interaction is described by
\begin{equation}\begin{split}&S^{(e-\omega)}=\overline{\psi}
(y)\Gamma^{(0)}_{\parallel}(y,x\mid \rho)\psi
(x)A_{\parallel}(\rho)\\&+\overline{\psi}
(y)\Gamma^{(0)}_{\perp}(y,x\mid \rho)\psi
(x)A_{\perp}(\rho).\nonumber\end{split}\end{equation} The
composite variables $y=\{\textbf{r},u\}$, $x=\{\textbf{r}',u'\}$,
and $\rho=\{\textbf{R},v\}$ are defined as follows:
$\textbf{r},\textbf{r}',\textbf{R}$ are 2D radius vectors, and
according to imaginary-time (Matsubara) formalism the variable
$u,u',v$ range from $0$ to $\hbar \beta=\hbar/(kT)$. We set
$\hbar=1$ and we use the summation-integration convention: that
repeated variables are summed up or integrated over.
 $G^{(0)-1}(y,x)$ is the inverse single-particle Green function for
non-interacting electrons in a periodic lattice potential
$G^{(0)-1}(y,x)=\sum_{\omega_{m}}e^{-\imath\omega_{m}(u-u')}
G^{(0)-1}(\textbf{r},z,\textbf{r}',z';\imath\omega_{m})$. The
function $G^{(0)-1}(\textbf{r},z,\textbf{r}',z';\imath\omega_{m})$
is defined as a sum of electron $\sum_{\textbf{k}_c}
\varphi^*_{c,\textbf{k}_c}(\textbf{r},z)\varphi_{c,\textbf{k}_c}(\textbf{r}',z')
G_{cc}^{(0)-1}(\textbf{k}_c;\imath\omega_{m})$ and hole
$\sum_{\textbf{k}_v}\varphi^*_{v,\textbf{k}_v}(\textbf{r},z)\varphi_{v,\textbf{k}_v}(\textbf{r}',z')
G_{vv}^{(0)-1}(\textbf{k}_v;\imath\omega_{m})$ parts. Here
$G_{cc}^{(0)-1}(\textbf{k}_c;\imath\omega_{m})=\imath\omega_{m}-\left[E_c(\textbf{k}_c,\lambda=1)-\mu_c\right]$,
and
$G_{vv}^{(0)-1}(\textbf{k}_v;\imath\omega_{m})=\imath\omega_{m}-\left[E_v(\textbf{k}_v,\xi=1)-\mu_v\right]$.
The functions $\varphi_{c,\textbf{k}_c}(\textbf{r})$ and
$\varphi_{v,\textbf{k}_v}(\textbf{r})$ are the wave functions of
the first electron and hole confined levels, defined by the
solutions of the corresponding Schrodinger equations. The electron
and hole chemical potentials are denoted by $\mu_c$ and $\mu_v$,
respectively, and the symbol $\sum_{\omega_{m}}$ is used to denote
$\beta^{-1}\sum_{m}$. For fermion fields we have
$\omega_{m}=(2\pi/ \beta)(m+1/2); m=0, \pm 1, \pm 2,...$. \\
In addition to the lattice potential, the electrons and holes
experience a Coulomb interaction, described by the term
$\Gamma_\parallel^{(0)}D^{(0)}_{\parallel}\Gamma_\parallel^{(0)}$:
\begin{equation}\begin{split}
&\Gamma_\parallel^{(0)}(y_2,x_1|\rho)D^{(0)}_{\parallel}(\rho,\rho')\Gamma_\parallel^{(0)}(y_3,x_4|\rho')=\\&\delta(u_1-u_3)
\delta(u_2-u_4)\sum_{\textbf{k}_i,\textbf{k}_j,\textbf{k}_i',\textbf{k}_j',\textbf{q}}\sum_{i,j}
\varphi_{i,\textbf{k}_i}(\textbf{r}_1)\varphi^*_{j,\textbf{k}_j}(\textbf{r}_2)\times
\\&
\varphi_{i,\textbf{k}'_{i}}(\textbf{r}_3)\varphi^*_{j,\textbf{k}'_{j}}(\textbf{r}_4)
V_0(\textbf{q})\left[\delta_{\textbf{k}_i,\textbf{k}'_{i}+\textbf{q}}+
\delta_{\textbf{k}_j,\textbf{k}'_{j}-\textbf{q}}\right].\nonumber\end{split}\end{equation}
Here ${i,j}=\{c,v\}$,  $D^{(0)}_{\parallel}$ is the longitudinal
part of the photon propagator (in a gauge, when the scalar
potential equals zero) and $\Gamma_\parallel^{(0)}$ is the vertex.
$V_0(\textbf{q})=2\pi e^2f(L|\textbf{q}|)/|\textbf{q}|$ denotes
the Fourier transform of the 2D unscreened Coulomb potential. The
structure factor $f(x)$ takes into account the first confined QW
electron and hole levels:
$$f(x)=\frac{3x^2+8\pi^2}{x(x^2+4\pi^2)}-\frac{32\pi^4[1-\exp(-x)]}{x^2(x^2+4\pi^2)^2}.$$ The inverse transverse photon
propagator is:
\begin{eqnarray}
D^{(0)-1}_{\perp}(\rho,\rho')=D^{(0)-1}_{\perp}(\textbf{R},v;\textbf{R}',v')=\nonumber\\\frac{1}{A_0}
\sum_{\textbf{q}}
\sum_{\omega_{p}}e^{\imath[\textbf{q.}(\textbf{R}-\textbf{R}')-
\omega_{p}(v-v')]}D^{(0)-1}_{\perp}(\textbf{q},\imath\omega_p)
.\label{PhGF0n}
\end{eqnarray}
  Here $A_0$ is the area of the cavity, and $D^{(0)-1}_{\perp}(\textbf{q},\imath\omega_p)=2\pi
c^2/[(\imath\omega_p)^2-\Omega^2(\textbf{q})]$. The symbol
$\sum_{\omega_{p}}$ is used to denote $\beta^{-1}\sum_{p}$. For
boson fields we have $\omega_{p}=(2\pi/ \beta)p ; p=0, \pm 1, \pm
2,...$ We assume that the electron-hole-photon system is in
thermal equilibrium, which means that the poles of the photon
retarded Green's function have to be obtained from the
corresponding Matsubara Green's function, by the substitution
$\imath\omega_p \rightarrow \omega+\mu+i0^+$. Here
$\mu=\mu_c-\mu_v$ is the chemical potential of the system. The
vertex $\Gamma^{(0)}_{\perp}(y,x\mid \rho)$ has the following
form:
\begin{equation}\begin{split}
&\Gamma^{(0)}_{\perp}(y_2,x_1\mid
\rho)=\Gamma^{(0)}_{\perp}(\textbf{r}_2,u_2,\textbf{r}_1,u_1\mid
\textbf{R},v)= \\&\frac{\delta(u_1-v)\delta(u_1-u_2)}{c}
  \sum_{\textbf{q}}\sum_{i,\textbf{k}_i,j,\textbf{k}_j}
e^{ \imath\textbf{q.R} }
\varphi^{*}_{j,\textbf{k}_j}(\textbf{r}_2)\varphi_{i,\textbf{k}_i}(\textbf{r}_1)\\&<j,\textbf{k}_j\mid
\widehat{\textbf{j}}(\textbf{q})\textbf{.n}(\textbf{q})\mid
i,\textbf{k}_i>,\label{Gamma0n}
\end{split}\end{equation}
where  $\widehat{\textbf{\textbf{j}}}(\textbf{q})$ denotes the
single-particle current operator, and
$\textbf{n}(\textbf{q})=\textbf{e}_\textbf{q}\times \textbf{e}_z$,
where $\textbf{e}_\textbf{q}=\textbf{q}/q$ and
$\textbf{e}_z=(0,0,1)$.\\It is well-known that all Green's functions
can be obtained by functional differentiation from the generating
functional $W[J,M]=\ln Z[J,M]$ of the connected Green's function,
where\begin{equation}\begin{split}&Z[J,M]= \int
D\mu[\overline{\psi},\psi,A_{\perp},A_{\parallel}]\times
\\&\exp[S+J_\perp(\rho) A_\perp(\rho)+J_\parallel(\rho)
A_\parallel(\rho)-\overline{\psi}(y)M(y,x)\psi(x)].\label{GF}\end{split}\end{equation}
Here $J$ and $M$ are the sources of the corresponding fields. By
means of the functional (\ref{GF}) we introduce the following
functions (after the functional differentiation one should set J=M=0):\\
Average photon field: \begin{equation}
R_{\parallel,(\perp)}(\rho)=-\frac{\delta W}{\delta
J_{\parallel(\perp)}(\rho)};\label{avPt}\end{equation}single-particle
Green function:
\begin{equation}
G(x,y)=-\frac{\delta W[J,M]}{\delta M(y,x)} ; \label{EGFt}
\end{equation} transverse
(longitudinal) photon Green function:
\begin{equation}D_{\perp(\parallel)}(\rho,\rho')=-\frac{\delta^2W[J,M]}{\delta
J_{\perp(\parallel)}(\rho')\delta
J_{\perp(\parallel)}(\rho)}=\frac{\delta
R_{\parallel,\perp}(\rho)}{\delta J_{\parallel,\perp}(\rho')};
\label{PGFt}
\end{equation}

two-particle electron-hole Green function:
\begin{equation}
K\left(%
\begin{array}{cc}
  x & y'  \\
  y & x' \\
\end{array}%
\right)=-\frac{\delta^2 W[J,M]}{\delta M(y',x')\delta
M(y,x)}=\frac{\delta G(x,y)}{\delta M(y',x')} ; \label{TGFt}
\end{equation}
transverse  (longitudinal) electron-photon vertex function:
\begin{equation}
\Gamma_{\perp(\parallel)}(y,x \mid \rho)=-\frac{\delta
G^{-1}(y,x)}{\delta J_{\perp(\parallel)}\rho')}
D^{-1}_{\perp(\parallel)}(\rho',\rho) . \label{VFt}
\end{equation}
As a consequence of the fact that the measure is invariant under
the translations $\overline{\psi}\rightarrow
\overline{\psi}+\delta\overline{\psi}$,
$A_{\perp,_\parallel}\rightarrow A_{\perp,_\parallel}+\delta
A_{\perp,_\parallel}$ we derive the SD equations: \cite{Z}
\begin{equation}\begin{split}
&J_{\parallel(\perp)}(\rho)-D^{(0)-1}_{\parallel(\perp)}(\rho,\rho')R_{\parallel(\perp)}(\rho')\\&+
\Gamma^{(0)}_{\parallel(\perp)}(y,x\mid
\rho)G(x,y)=0,\label{SE1n}\end{split}
\end{equation}
\begin{equation}
G^{-1}(y,x)-G^{(0)-1}(y,x)+M(y,x)+\Sigma(y,x)=0. \label{SE2n}
\end{equation}
The mass operator $\Sigma $ has the form:
\begin{eqnarray} &\Sigma(y,x)=\Gamma^{(0)}_{\perp}(y,x\mid
\rho)R_{\perp}(\rho)+\Gamma^{(0)}_{\parallel}(y,x\mid
\rho)R_{\parallel}(\rho)\nonumber\\&-\Gamma^{(0)}_{\parallel}(y,x'\mid
\rho)G(x',y')\Gamma_{\parallel}(y',x\mid \rho')
D_{\parallel}(\rho,\rho')\nonumber\\&-\Gamma^{(0)}_{\perp}(y,x'\mid
\rho)G(x',y')\Gamma_{\perp}(y',x\mid \rho') D_{\perp}(\rho,\rho').
\label{MO}
\end{eqnarray}
The charge neutrality leads to the photon field
$R_\parallel(\rho)$ that vanishes identically. Thus, all terms
proportional to $R_\parallel(\rho)$ (or
$D_\parallel(\rho,\rho')\Gamma^{(0)}_{\parallel}(y,x\mid
\rho')G(x,y)$) does not need to be taken into account because of
the global neutrality of the electron-hole system. In what follows
we assume the so-called Hartree-Fock approximation in which the
mass-operator has the form
\begin{eqnarray}&\Sigma(y,x)=\Gamma^{(0)}_{\perp}(y,x\mid
\rho)R_{\perp}(\rho)-\nonumber\\&\Gamma^{(0)}_{\parallel}(y,x'\mid
\rho)G(x',y')\Gamma_{\parallel}(y',x\mid \rho')
D_{\parallel}(\rho,\rho').\label{ZGK}\end{eqnarray} The first and
the second terms in (\ref{ZGK}) are called the Hartree term and
the Fock term, respectively. The presence of Bose-condensed
polaritons modifies the single-particle Green's functions, and
therefore one has to consider the so-called normal
$G^{-1}_{cc}(\textbf{k};\imath\omega_m)=G^{(0)-1}_{cc}(\textbf{k};\imath\omega_m)-\Sigma_{cc}(\textbf{k})$
and
$G^{-1}_{vv}(\textbf{k};\imath\omega_m)=G^{(0)-1}_{vv}(\textbf{k};\imath\omega_m)-\Sigma_{vv}(\textbf{k})$,
and anomalous
$G^{-1}_{cv}(\textbf{k};\imath\omega_m)=G^{-1}_{vc}(\textbf{k};\imath\omega_m)=-\Delta(\textbf{k})$
single-particle Green's functions (the spin degrees of freedom are
not included). The diagonal parts of the mass operator in the
Hartree-Fock approximation are as follows:
$\Sigma_{cc}(\textbf{k})=\sum_{\textbf{q}}V(\textbf{k}-\textbf{q})\sum_{\omega_m}G_{cc}(\textbf{q};\imath\omega_m)$,
$\Sigma_{vv}(\textbf{k})=\sum_{\textbf{q}}V(\textbf{k}-\textbf{q})\sum_{\omega_m}G_{vv}(\textbf{q};\imath\omega_m)$.
 Here $V(\textbf{q})=2\pi e^2f(L|\textbf{q}|)/\epsilon_\infty|\textbf{q}|$
denotes the screened Coulomb potential. Using (\ref{ZGK}) we
calculate for the non-diagonal parts of the mass operator in the
Hartree-Fock approximation:
\begin{equation}\Delta(\textbf{k})=\sum_{\textbf{q}}\Gamma^{(0)}_\perp(\textbf{q},\textbf{k})R_\perp(\textbf{q})
+\Delta_{exc}(\textbf{k}),\label{D}\end{equation} where $\Delta$
is the order parameter for the system. The first (Hartree) and the
second (Fock) terms in (\ref{D}) represent the photonic and the
excitonic contributions to the order parameter, respectively. The
exact form of $\Gamma^{(0)}_\perp(\textbf{q},\textbf{k})$ can be
calculated by means of (\ref{PhGF0n}) and (\ref{Gamma0n}), but we
assume that the photons are coupled to the electron-hole system
through the local interaction, i.e.
$\Gamma^{(0)}_\perp(\textbf{q},\textbf{k})=g(\textbf{q}-\textbf{k})$.
In this approximation the  photon field $R_\perp$ and the
excitonic  order $\Delta_{exc}$ parameter are defined as follows:
\begin{equation}R_\perp(\textbf{k})=\sum_{\textbf{q}}g(\textbf{k}-\textbf{q})\sum_{\omega_m}G_{cv}(\textbf{q};\imath\omega_m)/\left(\Omega^2(\textbf{k})-\mu^2\right),
\label{Ord1}\end{equation}
 \begin{equation}\Delta_{exc}(\textbf{k})=\sum_{\textbf{q}}V(\textbf{k}-\textbf{q})
 \sum_{\omega_m}G_{cv}(\textbf{q};\imath\omega_m).\label{Ord2}\end{equation}
The photonic and the excitonic order parameters are not
independent.
 Using Eq. (\ref{Ord2}) for the excitonic order parameter, we calculate
$\sum_{\omega_m}G_{cv}(\textbf{q};\imath\omega_m)=\sum_{\textbf{k}}V^{-1}(\textbf{q}-\textbf{k})\Delta_{exc}(\textbf{k})$,
and therefore, we obtain the following constraint:
\begin{equation}
R_\perp(\textbf{k})=\frac{1}{\Omega(\textbf{k})^2-\mu^2}\sum_{\textbf{q},\textbf{p}}
g(\textbf{k}-\textbf{q})V^{-1}(\textbf{q}-\textbf{p})\Delta_{exc}(\textbf{p}),\label{CN}
\end{equation} where we have introduced a function
$V^{-1}$ defined by $\sum_{\textbf{p}}
V(\textbf{k}-\textbf{p})V^{-1}(\textbf{p}-\textbf{q})=\delta(\textbf{k}-\textbf{q})$.
\\
In the Hartree-Fock approximation the normal single-particle Green's
functions are: \cite{Z}
\begin{equation}\begin{split}&G_{cc}(\textbf{k};\imath\omega_m)=\left[\frac{u_{\textbf{k}}^2}{\imath\omega_m-\omega_{+}(\textbf{k})}
+\frac{v_{\textbf{k}}^2}{\imath\omega_m-\omega_{-}(\textbf{k})}\right],\\
&G_{vv}(\textbf{k};\imath\omega_m))=\left[\frac{v_{\textbf{k}}^2}{\imath\omega_m-\omega_{+}(\textbf{k})}
+\frac{u_{\textbf{k}}^2}{\imath\omega_m-\omega_{-}(\textbf{k})}\right],\\
&G_{cv}(\textbf{k};\imath\omega_m)=G_{vc}(\textbf{k};\imath\omega_m)=\\&
u_{\textbf{k}}v_{\textbf{k}}\left[\frac{1}{\imath\omega_m-\omega_{+}(\textbf{k})}
-\frac{1}{\imath\omega_m-\omega_{-}(\textbf{k})}\right].\label{GFT}
\end{split}\end{equation}
Here, the following notations have been used:
\begin{equation}\begin{split}
&u^2(\textbf{k})=\frac{1}{2}\left[1+\frac{\eta(\textbf{k})}{\varepsilon(\textbf{k})}\right],
v^2(\textbf{k})=\frac{1}{2}\left[1-\frac{\eta(\textbf{k})}
{\varepsilon(\textbf{k})}\right],\\
&\varepsilon(\textbf{k})=\sqrt{\eta^2
(\textbf{k})+\Delta^2(\textbf{k})}\quad,\quad
\omega_{\pm}(\textbf{k})=\zeta(\textbf{k}) \pm
\varepsilon(\textbf{k}),\\
&\zeta(\textbf{k})=\frac{1}{2}\left[E_c(\textbf{k},1)+
E_v(\textbf{k},1)
-\mu_c-\mu_v\right]\\&+\frac{1}{2}\sum_\textbf{q}V(\textbf{k}-\textbf{q})\left[n_-(\textbf{q})-n_+(\textbf{q})\right],\nonumber\\
&\eta(\textbf{k})=\frac{1}{2}\left[E_c(\textbf{k},1)-
E_v(\textbf{k},1)-\mu \right]\\&
-\frac{1}{2}\sum_\textbf{q}V(\textbf{k}-\textbf{q})\left[1-\left[1-n_+(\textbf{q})-n_-(\textbf{q})\right]
\frac{\eta(\textbf{q})}{\varepsilon(\textbf{q})}\right],\label{N2}
\end{split}\end{equation}
where  $n_\pm(\textbf{k})=\left[1+\exp\left(\pm\beta
\omega_\pm(\textbf{k})\right)\right]^{-1}$. Below the critical
temperature the order parameter is developed
($\Delta(\textbf{k})\neq 0$) and the single-particle excitations are
coherent combinations of electron-like $\omega_{+}(\textbf{k})$ and
hole-like $\omega_{-}(\textbf{k})$ excitations, renormalized due to
the interaction with the cavity modes. The coefficients
$u(\textbf{k})$ and $v(\textbf{k})$ which are called coherent
factors, give the probability amplitudes of these states in the
actual mixture. \\
 It is expected that the BEC phenomenon is not sensitive to the difference in electron and hole effective
masses, and therefore,
 we assume
$m_c=m_v=2m_{exc}$, where $m_{exc}^{-1}=m_c^{-1}+m_v^{-1}$ is the
exciton reduced mass. The equal mass assumption simplifies very
much the calculations because in this case $\mu_c+\mu_v=E_g$,
$\zeta(\textbf{k})=0$,
$n_+(\textbf{k})=n_-(\textbf{k})=\left[1+\exp\left(\beta
\varepsilon (\textbf{k})\right)\right]^{-1}$, and
$\eta(\textbf{k})$ is defined by:
\begin{equation}\begin{split}
&\eta(\textbf{k})=\frac{1}{2}\left[\textit{E}_g+\frac{\pi^2}{2mL^2}
+\frac{\textbf{k}^2}{2m_{exc}}-\mu\right]\\&-\frac{1}{2}
\sum_{\textbf{q}}V(\textbf{k}-\textbf{q})\left[1-\tanh\left(\frac{\beta\varepsilon(\textbf{q})}{2}
\right)\frac{\eta(\textbf{q})} {\varepsilon
(\textbf{q})}\right].\label{eta}\end{split}\end{equation} Note,
that in our approach $\eta(\textbf{k})$ is defined
self-consistently  by the solution of Eq. (\ref{eta}), while MSL
neglected the contributions due to the diagonal parts
($\Sigma_{cc}(\textbf{k})$ and $\Sigma_{vv}(\textbf{k})$) of the
mass operator, using the following expression:
\begin{equation}\eta(\textbf{k})=\textbf{k}^2/4m_{exc}-\varepsilon_F,\label{MSL}\end{equation} where the
effective Fermi energy is defined by
$\varepsilon_F=p^2_F/4m_{exc}=(\mu- E_g-\pi^2/2m_{exc}L^2)/2$.  In
the high-density limit, the neglected diagonal parts of the mass
operator are responsible only for small renormalization of the
single-particle excitations $\omega_{\pm}(\textbf{k})$. But, in
the low-density limit the
diagonal parts of the mass operator are crucial, and therefore, they cannot be neglected.\\
The order parameter (\ref{D}) includes both excitonic and photonic
contributions:
$$\Delta(\textbf{k})=\sum_{\textbf{q}}g(\textbf{k}-\textbf{q})
R_\perp(\textbf{q})+\Delta_{exc}(\textbf{k}),$$ and is determined
by the constraint (\ref{CN}) and the following BCS self-consistent
equation for the excitonic order parameter:
\begin{equation}\Delta_{exc}(\textbf{k})=
\sum_{\textbf{q}}V(\textbf{k}-\textbf{q})\sum_{\omega_m}
\frac{\Delta(\textbf{q})}{\omega_m^2+\varepsilon^2
(\textbf{q})}.\label{BCS2} \end{equation} To make contact with the
equations of MSL in the absence of a disorder, we assume a
high-density limit. In the high-density regime the screened Coulomb
interaction could be replaced by a short-range contact interaction
with a coupling strength $g_c$ given by angular average over the
Fermi surface. Assuming that: (i) $g(\textbf{q})=g$, and (ii) the
order parameters are space independent
($R_\perp(\textbf{k})=R\delta_{\textbf{k},0}$, and
$\Delta_{exc}(\textbf{k})=\Delta_{exc}\delta_{\textbf{k},0}$), we
obtain the total order parameter $\Delta=gR+\Delta_{exc}$, where
$$\Delta_{exc}=g_c\sum_{\textbf{k}}\sum_{\omega_m}\frac{\Delta}{\omega_m^2+\varepsilon^2
(\textbf{k})}.$$ While the chemical potential $\mu$ does not exceed
the cavity-mode energy $\omega_c$ at $\textbf{k}=0$, the constrain
(\ref{CN}) assumes the form:
\begin{equation}R=\frac{g}{g_c\left(\omega_c^2-\mu^2\right)}\Delta_{exc}.\label{CN1}\end{equation}
 Similar relationship has been found by MSL, but with $\omega_c-\mu$ in place of
$\omega_c^2-\mu^2$. This is because of the different photon
Green's functions, used by MSL. In the high-density regime
$\eta(\textbf{k})=\textbf{k}^2/m_{exc}-\varepsilon_F$, and Eq.
 (\ref{BCS2}) assumes the form of the BCS gap equation for a
superconductor:
\begin{equation}1=\left(g_c+\frac{g^2}{\omega_c^2-\mu^2}\right)\sum_{\textbf{k}}\frac{1}{2\varepsilon(\textbf{k})}\tanh\left(\frac{\beta\varepsilon(\textbf{k})}{2}\right).
\label{BCSHD}\end{equation}
 When $g=0$, the last equation assumes the form (\ref{Zeqs}) with $\alpha=0$. \\ The chemical potential $\mu$ is a nontrivial function of the total number of
excitations $N=N_{ph}+N_{exc}$ in the condensate and should be
calculated by solving the BCS equation self-consistently. The
total number of photons is
$N_{ph}=\sum_{\textbf{k}}R^2_\perp(\textbf{k})$. The number of
condensed electron-hole pairs $N_{exc}$ is:
\begin{equation}
N_{exc}=\sum_{\textbf{k}}\left[1-\tanh\left(\frac{\beta\varepsilon(\textbf{k})}{2}
\right)\frac{\eta(\textbf{k})} {\varepsilon
(\textbf{k})}\right].\label{BCS4}\end{equation} Our next step is
to show that the photon Green's function and the two-particle
electron-hole Green's function have common poles - the excitation
spectrum in the presence of a condensed phase. To prove the last
we introduce a Legendre transform
\begin{equation}V[R,G]=W[J,M]+J_\alpha(\rho)R_\alpha(\rho)+M(y,x)G(x,y).
\label{LTn}\end{equation} The repeated Greek index $\alpha$ denotes
summation over the parallel $\parallel$ and the perpendicular
$\perp$ components of the corresponding quantity. By means of the
above Legendre transform, we derive the following exact
equations:\cite{Z}
\begin{equation}\begin{split}
&D_{\perp}(\rho,\rho')=D^{(0)}_{\perp}((\rho,\rho')+D^{(0)}_{\perp}(\rho,\rho'')\Gamma^{(0)}_{\perp}((y,x\mid
\rho'')\\&K\left(%
\begin{array}{cc}
  x & y'  \\
  y & x' \\
\end{array}%
\right)\Gamma^{(0)}_{\perp}(y',x'\mid
\rho''')D^{(0)}_{\perp}(\rho''',\rho') ,\label{EQ1t}
\end{split}\end{equation}
\begin{equation}\begin{split}
&\frac{\delta G(x,y)}{\delta
J_{\perp}(\rho)}=\\&K^{(0)}\left(%
\begin{array}{cc}
  x & y'  \\
  y & x' \\
\end{array}%
\right)\Gamma_{\perp}(y',x'\mid \rho')D_{\perp} (\rho',\rho)=\\&
K\left(%
\begin{array}{cc}
  x & y'  \\
  y & x' \\
\end{array}%
\right)\Gamma^{(0)}_{\perp}(y',x'\mid \rho')D^{(0)}_{\perp}
(\rho',\rho) ,\label{EQ2t}
\end{split}\end{equation}
\begin{equation}K^{-1}\left(%
\begin{array}{cc}
  y & x'  \\
  x & y' \\
\end{array}%
\right)=K^{(0)-1}\left(%
\begin{array}{cc}
  y & x'  \\
  x & y' \\
\end{array}%
\right)- I\left(%
\begin{array}{cc}
   y & x'  \\
  x & y' \\
\end{array}%
\right) ,\label{EQ3t}
\end{equation}
where $K^{(0)}\left(%
\begin{array}{cc}
  x & y'  \\
  y & x' \\
\end{array}%
\right)=G(x,y')G(x',y)$ is the free two-particle propagator and
the kernel of the BS equation (\ref{EQ3t}) is given by:
\begin{equation}\begin{split}&I\left(%
\begin{array}{cc}
  y & x'  \\
  x & y' \\
\end{array}%
\right)=\frac{\delta \Sigma (y,x)}{\delta
M(y',x')}\\&+\Gamma^{(0)}_{\alpha}(y,x\mid
\rho)D^{(0)}_{\alpha}(\rho,\rho') \Gamma^{(0)}_{\alpha}(y',x'\mid
\rho').\label{EQ4t}
\end{split}\end{equation}
There are two important conclusions that could be extracted from the
above equations. The first one follows from Eq. (\ref{EQ2t}). This
equation clearly indicates that the Fock term in Eq. (\ref{ZGK}) is
related to the two-particle Green's function, so we can write the
mass operator in the following form:
\begin{equation}\begin{split}
&\Sigma(y,x)=\Gamma^{(0)}_{\alpha}(y,x\mid
\rho)R_{\alpha}(\rho)\\&-\Gamma^{(0)}_{\alpha}(y,x'\mid
\rho)K\left(%
\begin{array}{cc}
  x' & y''  \\
  y' & x'' \\
\end{array}%
\right)\\&\Gamma^{(0)}_{\alpha}(y'',x''\mid \rho')D^{(0)}_{\alpha}
(\rho,\rho')G^{-1}(y',x). \label{MO1}\end{split}
\end{equation}
 The second conclusion follows from Eq. (\ref{EQ1t}). Obviously, the transverse photon Green's function
and the two-particle Green's function have common poles - cavity
polaritons. The poles of the transverse photon Green's function are
defined by the solutions of the Maxwell equations for a transverse
wave $\epsilon(\textbf{Q},\omega)=\Omega^2(\textbf{Q})/\omega^2$,
where the dielectric function $\epsilon(\textbf{Q},\omega)$  in the
case of a single excitonic resonance at energy $E_{l}(\textbf{Q})$
is:
$$\epsilon(\textbf{Q},\omega)=\epsilon_b+\frac{d}{E^2_{l}(\textbf{Q})-\omega^2-2\imath\omega\gamma_0}.$$
Here $\epsilon_b$ is the background dielectric constant,
$\gamma_0$ is the broadening of the excitonic resonance, and $d$
is proportional to the corresponding oscillator strength. In
principle, the excitonic resonance energy and the oscillator
strength can be calculated by solving a set of coupled BS
equations for the energy and the wavefunctions of the quantum-well
excitons. These equations are similar to equations (42) and (43)
from our earlier paper.\cite{Z} The only change that should be
done is related to the existence of an extra term,
$\Gamma^{(0)}_\perp D^{(0)}_\perp \Gamma^{(0)}_\perp G$, in the
mass operator. The last term contributes to the exchange
interaction in the similar manner as the term
$\Gamma^{(0)}_\parallel D^{(0)}_\parallel \Gamma^{(0)}_\parallel
G$ generates the analytical exchange interaction between electrons
and holes. The exchange interactions are important only for the
fine structure of exciton levels, and therefore, we neglect the
exchange interaction terms. As a result, we obtain the BS
equations similar to the case of the excitonic condensate.
Generally speaking, to calculate the excitation spectrum in the
presence of a condensed phase, one has to solve the BCS and the BS
equations simultaneously, taking into account the fact that the
chemical potential depends on the number of excitons and photons
in the condensed phase. Such an ambitious
task will be left as a subject of future research.\\
 We finish this section with a brief
discussion of the so-called Thouless criterion. \cite{Th}  In the
case of superconductivity this criterion says that below the
critical temperature $T_c$ the T-matrix has a pole at zero
frequency and zero momentum (the existence of Goldstone mode below
$T_c$). To check whether the Thouless criterion works in the case
of condensed cavity polaritons we follow the method based on the
Ward identities.\cite{RH} Taking into account the fact that in the
case of MC polaritons the mass operator depends on the photon
field $R_{\perp(\parallel)}$ and the Green's function $G$, we
first invert the SD equations to express the sources
$J_\alpha(\rho)$ and $M(y,x)$ as functionals of the field
$R_\alpha$ and the Green's function $G$. Second, we assume that
there exists a continuous transformation, for example, a rotation
in order-parameter space, which depends continuously on the
parameter $\lambda$. If the system is invariant under this
transformation, then the variation of the Legendre transform
implied by the transformation is equal to zero, i.e.
$\delta_\lambda V=0$. The BS equation for the two-particle Green's
function $K=K^{(0)}+K^{(0)}IK$ can be rewritten in terms of the
many-particle T-matrix, $T=I+IK^{(0)}T$, in the form
$K=K^{(0)}+K^{(0)}TK^{(0)}$. By means of the last form of the BS
equation and the definition (\ref{TGFt}) we calculate the
variation of the inverse Green's function $\delta_\lambda
G^{-1}=-\left(1+TK^{(0)}\right)\delta_\lambda M$. Using the SD
equations we find $\delta_\lambda G^{-1}=-\delta_\lambda
M-\delta_\lambda \Sigma$, and therefore, $T^{-1}\delta_\lambda
\Sigma=K^{(0)}\delta_\lambda M$. But, according to (\ref{LTn}) we
calculate $\delta_\lambda M=\frac{\delta}{\delta
G}\left(\delta_\lambda V\right)=0$, and therefore,
$T^{-1}\delta_\lambda \Sigma=0$. Above the critical temperature
$T_c$ the order parameter is zero, and hence, $\delta_\lambda
\Sigma=0$. Thus,  $T^{-1}\delta_\lambda \Sigma=0$ is satisfied
trivially. Below $T_c$ the order parameter is nonzero and
$\delta_\lambda \Sigma\neq 0$, which requires that the inverse
T-matrix has a zero eigenvalue. Thus, we conclude that below the
critical temperature
 $T_c$, the T-matrix  must have a pole at zero energy and zero momentum.
 The existence of Goldstone mode below $T_c$ indicates that the
 formation of a condensate in MC is possible.

\section{\label{sec:sec3}Cavity polaritons in the presence of
a symmetry-breaking disorder - the Keldysh Green's function
approach}

In the presence of a disorder we define the Green's functions as
an average of the time-ordered products of the fields, but the
average includes both the quantum and the disorder averaging.  We
use $<F>$ for the quantum averaging, and $\overline{F}$ for the
disorder averaging. The Matsubara Green's function approach
discussed in the previous Section, cannot be directly applied to
the cavity polaritons in the presence of a disorder, because one
has to calculate $\overline{<\ln F>}$. Nevertheless, for a given
disorder configuration we can write the relationship between the
random photon Green's function $G$ and the random two-particle
Green's function $K$, similar to Eq. (\ref{EQ1t}). After that, the
disorder averaging replaces the random Green's functions on the
both sides of Eq. (\ref{EQ1t}) by their averages. Thus, we obtain
that average photon $\overline{D}$ and average two-particle
Green's function $\overline{K}$ have common poles - the cavity
polaritons in the presence of a disorder. The next question to be
answered is about the possibility to observe a condensate in MC in
the presence of a disorder. To answer this question we have to
examine the non-diagonal parts of the average single-particle
Green's function. The last can be obtained by solving the SD
equations. We shall use the Keldysh technique to derive the SD
equations in the presence of a disorder, because the CPT approach
allows us to perform the average over the random potential
exactly, i.e. non-perturbatively. Within this approach we have
boson (photon) longitudinal and transverse fields $A_{\alpha}(
\underline{z})=A_{\parallel,\perp}(\textbf{R}, \underline{t''})$
interacting with a fermion (electron) field $\psi^+
(\underline{y})=\psi_j^+ (\textbf{r},\underline{t})$, or
$\psi(\underline{x})=\psi_{j'} (\textbf{r}',\underline{t}')$, in
the presence of disorder. The variables $\underline{y},
\underline{x}$ and $\underline{z}$ are composite variables
$\underline{y}=\{\textbf{r}, \underline{t},j\}$,
$\underline{x}=\{\textbf{r}',\underline{t}',j'\}$,
$\underline{z}=\{\textbf{R},\underline{t}''\}$, where
$\textbf{r},\textbf{r}',\textbf{R}$ are the corresponding 2D
radius vectors. The index $j=1$ (or $j=c$) denotes the electron
states, and $j=2$ (or $j=v$) denotes the hole states. For
simplicity, we suppose that the electrons are spinless. At a zero
temperature the total action of the system is $S=
S^{(e)}+S^{(\omega)}_0+S^{(e-\omega)}$. To incorporate the effect
of disorder we introduce random static symmetry-breaking
(charge-dependent) potential $V(\textbf{r})$. In the presence of a
disorder the action which corresponds to the electron system
assumes the form:
$$S^{(e)}=\psi^+
(\underline{y})[G^{(0)-1}(\underline{y},\underline{x})-V(\underline{y},\underline{x})]\psi(\underline{x}),$$
where
$V(\underline{y},\underline{x})=V(\textbf{r},\underline{t},j;\textbf{r}',\underline{t}',j')=
V(\textbf{r})
\delta(\textbf{r}-\textbf{r}')\delta(\underline{t}-\underline{t}')\delta_{jj'}.$
  The actual potential
$V(\textbf{r})$ is unknown, but we assume that it obeys Gaussian
statistics such that:
\begin{equation}\overline{V(\textbf{r})}=0,\quad
\overline{V(\textbf{r})V(\textbf{r}')}=\Lambda\delta(\textbf{r}-\textbf{r}'),\label{RP1}\end{equation}
where $\Lambda=(2\pi \nu \tau)^{-1}$ , $\nu=m_{exc}/\pi$ is the 2D
density of states, and $\tau$ is the corresponding scattering
time. The disorder averaging in (\ref{RP1}) is defined for any
functional $F[V]$ by the functional integral:
\begin{equation}\overline{F}= \int DVF[V]\\
\exp[-\frac{1}{2\Lambda}\int V^2(\textbf{r})d\textbf{r}].
\label{RPI}\end{equation} The actions $S^{(\omega)}_0$ and
$S^{(e-\omega)}$ are given by:
$$S^{(\omega)}_0=\frac{1}{2}A_{\alpha}(\underline{z})D^{(0)-1}_{\alpha}(\underline{z},\underline{z}')A_{\alpha}(\underline{z}'),$$
$$S^{(e-\omega)}=\psi^+
(\underline{y})\Gamma^{(0)}_{\alpha}(\underline{y},\underline{x}\mid
\underline{z})\psi(\underline{x})A_{\alpha}(\underline{z}).$$ In the
CPT formalism, the Green's functions are defined by means of two
time orderings in the same formula. In what follows we use the
single-time representation. In this representation the two time
orderings are replaced by a single time ordering along the Keldysh
contour which at a zero temperature consists of two branches: the
right-going $(+)$ from $-\infty$ to $\infty$ and the left-going
$(-)$ from $\infty$ to $-\infty$. The symbol $\underline{t}$ means
that the time integral $\int d\underline{t}$ along the Keldysh
contour could be written as two usual integrals, i.e. $\int
d\underline{t}=\int_{-\infty}^\infty dt^+ -\int_{-\infty}^{\infty
}dt^-$. In other words, the time variable
 $\underline{t}$  on the positive branch equals $\underline{t}=t^+$, and
$\underline{t}=t^-$ on the negative branch.  It should be
mentioned that our equations are valid for both nonequilibrium and
equilibrium conditions, but we do not discuss time-dependent
phenomena on an ultrafast scale. Instead, we intend to describe
steady-state phenomenon, such as the light propagation in crystal
in terms of excitonic polaritons. In the steady-state regime all
quantities depend on the relative time
$\underline{t}'-\underline{t}''$. \\The inverse free propagator
$G^{(0)-1}(\underline{y},\underline{x})=G^{(0)-1}_{jj'}(\textbf{r},\underline{t};\textbf{r}',\underline{t}')$
is defined as follows:
\begin{widetext}
\begin{equation}\begin{split}
&G^{(0)-1}_{jj'}(\textbf{r},\underline{t};\textbf{r}',\underline{t}')=
\delta(\underline{t}-\underline{t}')\delta_{jj'}\delta_{jc}\sum_{\textbf{k}_c}\varphi_{c,\textbf{k}_c}(\textbf{r})\varphi^*_{c,\textbf{k}_c}(\textbf{r}')
\int_{-\infty}^{\infty}\frac{d \omega}{2\pi}
G_{c}^{(0)-1}(\textbf{k}_c;\omega)e^{\imath \omega \underline{t}}
\\&+\delta(\underline{t}-\underline{t}')\delta_{jj'}\delta_{jv}
\sum_{\textbf{k}_v}\varphi_{v,\textbf{k}_v}(\textbf{r})\varphi^*_{v,\textbf{k}_v}(\textbf{r}')
\int_{-\infty}^{\infty}\frac{d \omega}{2\pi}
G_{v}^{(0)-1}(\textbf{k}_v;\omega)e^{\imath \omega \underline{t}}
\label{ElGF} \end{split}\end{equation} Here
$G_{c}^{(0)-1}(\textbf{k}_c;\omega)=\omega-\left[E_c(\textbf{k}_c,\lambda=1)-\mu_c\right]+\imath
0^+$,
$G_{vv}^{(0)-1}(\textbf{k}_v;\omega)=\omega-\left[E_v(\textbf{k}_v,\xi=1)-\mu_v\right]-\imath
0^+$ are the inverse zero-temperature free electron and hole
propagators. \\
In addition to the lattice and the random potentials, the
electrons and holes experience a Coulomb interaction, described by
the term
$\Gamma_\parallel^{(0)}D^{(0)}_{\parallel}\Gamma_\parallel^{(0)}$:
\begin{equation}\begin{split}
&\Gamma_\parallel^{(0)}(\underline{y_2},\underline{x_1}|\underline{z})D^{(0)}_{\parallel}(\underline{z},\underline{z}')
\Gamma_\parallel^{(0)}(\underline{y_3},\underline{x_4}|\underline{z}')=\delta(\underline{t_1}-\underline{t_3})
\delta(\underline{t_2}-\underline{t_4})\sum_{j,j'}\sum_{\textbf{k}_j,\textbf{p}_{j'},\textbf{q}}V_0(\textbf{q})
\varphi_{j,\textbf{k}_j}(\textbf{r}_1)\varphi_{j,\textbf{k}_j+\textbf{q}}^*(\textbf{r}_2)\times\\&
\varphi^*_{j',\textbf{p}_{j'}}(\textbf{r}_3)
\varphi_{j',\textbf{p}_{j'}-\textbf{q}}(\textbf{r}_4) .
\end{split}\end{equation} Here $D^{(0)}_{\parallel}$ is the
longitudinal part of the photon propagator (in a gauge, when the
scalar potential equals zero) and $\Gamma_\parallel^{(0)}$ is the
vertex. $V_0(\textbf{q})$ denotes the Fourier transform of the 2D
bare Coulomb potential, and has been defined in Sec. II. \\ The
inverse transverse photon propagator is:
\begin{equation}
D^{(0)-1}_{\perp
}(\underline{z},\underline{z}')=D^{(0)-1}_{\perp}(\textbf{R},\underline{t};\textbf{R}',\underline{t}')=\frac{\delta(\underline{t}-\underline{t}')}{A}
\sum_{\textbf{q}} \int_{-\infty}^\infty \frac{d\omega}{2\pi}
e^{\imath[\textbf{q.}(\textbf{R}-\textbf{R}')-
\omega\underline{t}]}D^{(0)-1}_{\perp}(\textbf{q};\omega),\label{PhGF0}
\end{equation}
  Here $A$ is the area of the cavity, and $D^{(0)-1}_{\perp}(\textbf{q},\omega)=2\pi
c^2/[(\omega-\mu)^2-\Omega^2(\textbf{q})+\imath 0^+]$. The vertex
$\Gamma^{(0)}_{\perp}(\underline{y},\underline{x}\mid
\underline{z})$ has the following form:
\begin{equation}\begin{split}
&\Gamma^{(0)}_{\perp}(\underline{y_2},\underline{x_1}\mid
\underline{z})=\Gamma^{(0)}_{\perp
jj'}(\textbf{r}_2,\underline{t_2},\textbf{r}_1,\underline{t_1}\mid
\textbf{R},\underline{t})=\\&
\frac{\delta(\underline{t_1}-\underline{t})\delta(\underline{t_1}-\underline{t_2})}{c}
  \sum_{\textbf{q}}\sum_{\textbf{k}_{j},\textbf{p}_{j'}}
e^{ \imath\textbf{q.R} }
\varphi^{*}_{j,\textbf{k}_{j}}(\textbf{r}_2)\varphi_{j',\textbf{p}_{j'}}(\textbf{r}_1)<j,\textbf{k}_{j}\mid
\widehat{\textbf{j}}(\textbf{q})\textbf{.n}(\textbf{q})\mid
j',\textbf{p}_{j'}>,\label{Gamma0}
\end{split}\end{equation}\end{widetext}
where  $\widehat{\textbf{\textbf{j}}}(\textbf{q})$ denotes the
single-particle current operator, and
$\textbf{n}(\textbf{q})=\textbf{e}_\textbf{q}\times \textbf{e}_z$,
where $\textbf{e}_\textbf{q}=\textbf{q}/q$ and
$\textbf{e}_z=(0,0,1)$.\\
Let us introduce the generating functional $\textbf{W}[J,M;V]$  of
the connected Green functions:
\begin{equation}
\textbf{W}[J,M;V]=-\imath\textrm{ln} \textbf{Z}[J,M;V],\label{GF1}
\end{equation}
where $J=J_{\parallel,\perp}(\underline{z})$ and
$M=M(\underline{y},\underline{x})$ are the sources, and the
functional $\textbf{Z}[J,M;V]$ has the form:
\begin{equation}\begin{split}
&\textbf{Z}[J,M;V]=\int
D\mu\exp\{\imath[S\\&+J_{\parallel}(\underline{z})A_{\parallel}(\underline{z})+J_{\perp}(\underline{z})A_{\perp}(\underline{z})-
\psi^+(\underline{y})M(\underline{y},\underline{x})\psi
(\underline{x})]\}.\label{W}\end{split}
\end{equation}
 Here $D\mu= C D\psi^+ D\psi DA$ denotes the functional
measure. The success of the Keldysh technique is based on the fact
that the normalization constant $C$ is disorder-independent. Thus,
we assume that  $C$  is chosen in the manner that
$\textbf{Z}[J=0,M=0;V]=1$. It is clear that because  $J$ and $M$
do not have the same behavior on the forward
 and backward parts of the Keldysh contour, the generating functional
 is not equal to unity if the sources are not nullified. \\
By means of the generating functional of the connected Green
functions we introduce the following
average quantities:\\
Photon field $R_\alpha(\underline{z})$ (in what follows $\alpha=
\parallel$, or $\alpha=\perp$):
\begin{equation}R_\alpha(\underline{z})=\overline{\frac{\delta \textbf{W}[J,M;V]}{\delta
J_\alpha(\underline{z})}|_{J=M=0}}\quad;\label{AvPh}\end{equation}
 photon
Green's function $D_{\alpha}(\underline{z},\underline{z}')$:
\begin{equation}D_{\perp(\parallel)}(\underline{z},\underline{z}')=-\overline{\frac{\delta^2 \textbf{W}[J,M;V]}{\delta
J_{\perp(\parallel)}(\underline{z}) \delta
J_{\perp(\parallel)}(\underline{z}')}|_{J=M=0}}\quad; \label{PGF}
\end{equation} single-particle Green's function
$G(\underline{x},\underline{y})$:
\begin{equation}G(\underline{x},\underline{y})=-\imath\overline{\frac{\delta
\textbf{W}[J,M,V]}{\delta M(\underline{y},\underline{x})}|_{J=M=0}}
\quad; \label{EGF}
\end{equation} two-particle electron-hole Green's function $K\left(%
\begin{array}{cc}
\underline{x} & \underline{y}'  \\
  \underline{y} & \underline{x}' \\
\end{array}%
\right)$:
\begin{equation}K\left(%
\begin{array}{cc}
\underline{x} & \underline{y}'  \\
  \underline{y} & \underline{x}' \\
\end{array}%
\right)=-\overline{\frac{\delta^2 \textbf{W}[J,M;V]}{\delta
M(\underline{y},\underline{x})\delta
M(\underline{y}',\underline{x}')}|_{J=M=0}}. \label{TGF}
\end{equation}
Evidently, we have four single-electron Green's functions
$G^{\eta_1\eta_2}$ and four photon Green's functions
$D^{\eta_1\eta_2}$, where $\eta_1,\eta_2=+$ or $-$ depending on
whether the time variable $\underline{t}$ is on the positive
branch or on the negative branch. Note, that the time integration
in expressions like
$C(\underline{x}_1,\underline{x}_3)=A(\underline{x}_1,\underline{y}_2)B(\underline{y}_2,\underline{x}_3)$
 follow the convention:
\begin{equation}\begin{split}
&C^{\eta_1\eta_2}(\underline{x}_1,\underline{x}_3)=C(\textbf{r}_1,t_1^{\eta_1};\textbf{r}_3,t_3^{\eta_2})\\&=\int
d\textbf{r}_2\int_{-\infty}^\infty dt_2^+
A(\textbf{r}_1,t_1^{\eta_1};\textbf{r}_2,t_2^+)B(\textbf{r}_2,t_2^{+};\textbf{r}_3,t_3^{\eta_2})\\
&-\int d\textbf{r}_2\int_{-\infty}^\infty dt_2^-
A(\textbf{r}_1,t_1^{\eta_1};\textbf{r}_2,t_2^-)B(\textbf{r}_2,t_2^{-};\textbf{r}_3,t_3^{\eta_2}).\label{Conv}\end{split}\end{equation}
 Due to the time-translational
invariance $K\left(%
\begin{array}{cc}
 \textbf{r}_1,\underline{t}_1 & \textbf{r}_3,\underline{t}_3 \\
  \textbf{r}_2,\underline{t}_2 & \textbf{r}_4,\underline{t}_4 \\
\end{array}%
\right)$ depends on $\underline{t}_{12}=\underline{t}_1
-\underline{t}_2,\underline{t}_{43}= \underline{t}_4
-\underline{t}_3$ and $\underline{t}_{31} =\underline{t}_3
-\underline{t}_1 $. In what follows, we shall see that our
equations will involve the two-particle Green's functions with
$\underline{t}_{12}=\underline{t}_{43}=0$, and therefore, we have
four different two-particle Green's functions
$K^{\eta_1\eta_2}=K\left(%
\begin{array}{cc}
 \textbf{r}_1,\underline{t}_1^{\eta_1}& \textbf{r}_3, \underline{t}_3^{\eta_2}\\
  \textbf{r}_2,\underline{t}_1^{\eta_1} & \textbf{r}_4,\underline{t}_3^{\eta_2} \\
\end{array}%
\right)$. The corresponding retarded Green's functions
$G^R,D^R,K^R$, for example, $G^R$, can be expressed as
$G^R=G^{--}-G^{-+}=G^{+-}-G^{++}.$ \\
The Keldysh technique allows us to perform the disorder averaging.
The resulting equations for the average photon field
$R_\alpha(\underline{z})$ and the average single-particle Green's
function $G(\underline{x},\underline{y})$ are as follows:
\begin{equation}R_\alpha(\underline{z})=-\imath\frac{\delta Z[J,M]}{\delta
J_\alpha(\underline{z})}|_{J=M=0},\label{D1}\end{equation}
\begin{equation}G(\underline{x},\underline{y})=-
\frac{\delta Z[J,M]}{\delta
M(\underline{y},\underline{x})}|_{J=M=0}.\label{D2}
\end{equation}
\begin{widetext}
where the average generating functional
$Z[J,M]=\overline{\textbf{Z}[J,M;V]}$ is defined by the equation:
\begin{equation}\begin{split} &Z[J,M]=\int
D\mu\exp\{\imath[\psi^+
(\underline{y})G^{(0)-1}(\underline{y},\underline{x})\psi
(\underline{x})+\frac{1}{2}A_{\alpha}(\underline{z})D^{(0)-1}_{\alpha
}(\underline{z},\underline{z}')A_{\alpha}(\underline{z}')+\psi^+
(\underline{y})\Gamma^{(0)}_{\alpha}(\underline{y},\underline{x}\mid
\underline{z})\psi(\underline{x})A_{\alpha}(\underline{z})\\&+\frac{\imath}{2}\Lambda
\psi_j^+
(\textbf{r},\underline{t})\psi_j(\textbf{r},\underline{t})\psi_{j'}^+
(\textbf{r},\underline{t}')\psi_{j'}(\textbf{r},\underline{t}')
+J_{\alpha}(\underline{z})A_{\alpha}(\underline{z})-
\psi^+(\underline{y})M(\underline{y},\underline{x})\psi
(\underline{x})]\}.\label{ZV}\end{split}
\end{equation}
To calculate $R$ and $G$ one has to know the functional $Z[J,M]$.
Note that $Z[J=0,M=0]=\overline{\textbf{Z}[J=0,M=0;V]}=1$. \\
Let us define the generating functional $W[J,M]=-\imath \ln Z[J,M]$.
By means of this definition we introduce two new functionals:
$R_\alpha(\underline{z};J,M)=\delta W[J,M]/\delta
J_\alpha(\underline{z})$ and
$G(\underline{x},\underline{y};J,M)=-\imath \delta W[J,M]/\delta
M(y,x)$. When the sources are nullified, the new functionals are
equal to the average photon field $R_\alpha(\underline{z})$ and to
the average Green's function $G(\underline{x},\underline{y})$,
respectively:
$$R_\alpha(\underline{z};J,M)|_{J=M=0}=\frac{\delta W[J,M]}{\delta J_\alpha(\underline{z})}|_{J=M=0}=-\imath \frac{1}{Z[J,M]}|_{J=M=0}
\frac{\delta Z[J,M]}{\delta
J_\alpha(\underline{z})}|_{J=M=0}=-\imath \frac{\delta
Z[J,M]}{\delta
J_\alpha(\underline{z})}|_{J=M=0}=R_\alpha(\underline{z})$$
$$G(\underline{x},\underline{y};J,M)|_{J=M=0}=-\imath\frac{\delta W[J,M]}{\delta M(\underline{y},\underline{x})}|_{J=M=0}=-\frac{1}{Z[J,M]}|_{J=M=0}
\frac{\delta Z[J,M]}{\delta
M(\underline{y},\underline{x})}|_{J=M=0}=- \frac{\delta
Z[J,M]}{\delta
M(\underline{y},\underline{x})}|_{J=M=0}=G(\underline{x},\underline{y})$$
The next step is to derive the SD equations for the corresponding
average quantities using the fact that the functional measure in
(\ref{ZV}) is invariant under the translations $\psi^+\rightarrow
\psi^++\delta\psi^+$, $A\rightarrow A+\delta A$. This assumption
yields highly nontrivial relations among generating functionals and
their derivatives which are as follows:
\begin{equation}0=J_{\parallel(\perp)}(\underline{z})+
D^{(0)-1}_{\parallel(\perp)}(\underline{z},\underline{z}')
R_{\parallel(\perp)}(\underline{z}';J,M)-\imath
\Gamma^{(0)}_{\parallel(\perp)}(\underline{y},\underline{x}\mid
\underline{z})G(\underline{x},\underline{y};J,M), \label{SDE1V}
\end{equation}
\begin{equation}G^{-1}(\underline{y},\underline{x};J,M)=
G^{(0)-1}(\underline{y},\underline{x})-M(\underline{y},\underline{x})
-\Sigma(\underline{y},\underline{x};J,M). \label{SDE2V}
\end{equation}  Here,
$\Sigma(\underline{y},\underline{x};J,M)$ is a functional of the
sources, but when the sources are nullified we obtain the mass
operator $\Sigma(\underline{y},\underline{x})$ for the average
single-particle Green's function:
\begin{equation}\begin{split}
&\Sigma(y,x)=
-\Gamma^{(0)}_{\parallel}(\underline{y},\underline{x}'\mid
\underline{z})R_{\parallel}(\underline{z})-\Gamma^{(0)}_{\perp}(\underline{y},\underline{x}'\mid
\underline{z})R_{\perp}(\underline{z})
-\Gamma^{(0)}_{\parallel}(\underline{y},\underline{x}'\mid
\underline{z})\left(-\imath \frac{\delta
G(\underline{x}',\underline{y}';J,M)}{\delta
J_{\parallel}(\underline{z})}|_{J=M=0}\right)
G^{-1}(\underline{y}',\underline{x})\\
&-\Gamma^{(0)}_{\perp}(\underline{y},\underline{x}'\mid
\underline{z})\left(-\imath \frac{\delta
G(\underline{x}',\underline{y}';J,M)}{\delta
J_{\perp}(\underline{z})}|_{J=M=0}\right)
G^{-1}(\underline{y}',\underline{x})-
\imath\Lambda K\left(%
\begin{array}{cc}
\textbf{r},\underline{t},j & \textbf{r},\underline{t}^I,j^I   \\
  \textbf{r}^{II},\underline{t}^{II},j^{II}  & \textbf{r},\underline{t}^I,j^I  \\
\end{array}%
\right)G^{-1}_{j^{II}
j'}(\textbf{r}^{II},\underline{t}^{II},\textbf{r}',\underline{t}')
. \label{AvMOp1}\end{split}
\end{equation}
The term proportional to $R_\parallel(z)$ does not need to be
taken into account because of the charge neutrality of the system.\\
The next step is to find the relationship between the mass operator
and the average two-particle Green's function. By solving the SD
equations (\ref{SDE1V}) and (\ref{SDE2V}), one can obtain the
sources $J_\alpha$ and $M$ as functionals of $R_\alpha$ and $G$. By
means of the identity:
$$0=\frac{\delta J_\alpha(\underline{z})}{\delta M(\underline{y},\underline{x})}
=\frac{\delta J_\alpha(\underline{z})}{\delta
G(\underline{x}',\underline{y}')}\frac{\delta
G(\underline{x}',\underline{y}')}{\delta
M(\underline{y},\underline{x})}+\frac{\delta
J_\alpha(\underline{z})}{\delta R_\beta(\underline{z})}\frac{\delta
R_\beta(\underline{z})}{\delta M(\underline{y},\underline{x})},$$ we
calculate
\begin{equation}
-\imath \frac{\delta G(\underline{x}',\underline{y}';J,M)}{\delta
J_{\parallel(\perp)}(\underline{z})}|_{J=M=0}= -\imath \frac{\delta
G(\underline{x}',\underline{y}';J,M)}{\delta
M(\underline{y}',\underline{x}')}|_{J=M=0}\Gamma^{(0)}_{\parallel(\perp)}(\underline{y}',\underline{x}'\mid
\underline{z}')D^{(0)}_{\parallel(\perp)}(\underline{z}',\underline{z}).
\label{FF}\end{equation}
 In the absence of a symmetry-breaking disorder the $
 -\imath\delta G/\delta M$ is noting but the two-particle Green's function $K$. However, in the presence of a symmetry-breaking
 disorder
  the average generating functional $Z[J,M]$ is not enough
to obtain  the average photon Green's function and the average
two-particle Green's function, because they both can be written as
functional derivatives of $Z[J,M]$ plus terms not directly related
to the functional $Z[J,M]$ or its derivatives:
\begin{equation}\begin{split}
&K\left(%
\begin{array}{cc}
\underline{x} & \underline{y}'  \\
  \underline{y} & \underline{x}' \\
\end{array}%
\right)=\imath\frac{\delta^2Z[J,M]}{\delta
M(\underline{y},\underline{x})\delta
M(\underline{y}',\underline{x}')}|_{J=M=0}-\imath\overline{G(\underline{x},\underline{y};V,J,M)G(\underline{x}',\underline{y}';V,J,M)}|_{J=M=0}\\
&=-\imath \frac{\delta G(\underline{x},\underline{y};J,M)}{\delta
M(\underline{y}',\underline{x}')}|_{J=M=0}+\imath
G(\underline{x},\underline{y})G(\underline{x}',\underline{y}')-\imath\overline{G(\underline{x},\underline{y};V,J,M)G(\underline{x}',\underline{y}';V,J,M)}|_{J=M=0}
,\label{R1}
\end{split}\end{equation}
\begin{equation}\begin{split}
&D_{\perp(\parallel)}(\underline{z},\underline{z}')=\imath\frac{\delta^2Z[J,M]}{\delta
J_{\perp(\parallel)}(\underline{z})\delta
J_{\perp(\parallel)}(\underline{z}')}|_{J=M=0}+\imath\overline{R_{\perp(\parallel)}(\underline{z};V,J,M)R_{\perp(\parallel)}(\underline{z}';V,J,M)}|_{J=M=0}\\
&=-\frac{\delta R_{\perp(\parallel)}(\underline{z};J,M)}{\delta
J_{\perp(\parallel)}(\underline{z})}|_{J=M=0}-\imath
R_{\perp(\parallel)}(\underline{z})R_{\perp(\parallel)}(\underline{z}')
+\imath\overline{R_{\perp(\parallel)}(\underline{z};V,J,M)R_{\perp(\parallel)}(\underline{z}';V,J,M)}|_{J=M=0}
.\label{R3}\end{split}\end{equation} Here, we have introduced the
functionals:
\begin{equation}G(\underline{x},\underline{y};V,J,M)=-\imath
\frac{\delta \textbf{W}[J,M;V]} {\delta
M(\underline{y},\underline{x})},\quad
R_{\parallel(\perp)}(\underline{z};V,J,M)=\frac{\delta
\textbf{W}[J,M;V]}{\delta
J_{\parallel(\perp)}(\underline{z})}\label{2F}\end{equation} which
depend on the random potential $V$. Since the average of the
product $\overline{G.G}$ (or $\overline{R.R}$) is not the product
of the averages $\overline{G}.\overline{G}$ (or
$\overline{R}.\overline{R}$), the term $\overline{G.G}$ in
(\ref{R1}) does not cancel $\overline{G}.\overline{G}$. We have
already mentioned that by performing the disorder averaging of
both sides of Eq. (\ref{EQ1t})  we can obtain a relationship
between the average photon and the average two-particle Green's
functions. In other words, the average Green's functions
(\ref{R1}) and (\ref{R3}) must satisfy the equation
 $D_\perp=D_\perp^{(0)}+D_\perp^{(0)}\Gamma_\perp^{(0)}K\Gamma_\perp^{(0)}D_\perp^{(0)}$.
Obviously, from the SD equations (\ref{SDE1V}) and (\ref{SDE2V})
follows that the terms $\overline{G.G}$, $\overline{R.R}$,
$\overline{G}.\overline{G}$ and $\overline{R}.\overline{R}$ in
both sides of the last equation cancel each others. Thus, the
cavity polaritons in the presence of a symmetry-breaking disorder
manifest themselves as common poles of the term  $-\imath
\frac{\delta G(\underline{x},\underline{y};J,M)}{\delta
M(\underline{y}',\underline{x}')}|_{J=M=0}$ in (\ref{R1}) and
$-\frac{\delta R_{\perp}(\underline{z};J,M)}{\delta
J_{\perp}(\underline{z})}|_{J=M=0}$ in (\ref{R3}).\\
Next, we rewrite the mass operator $\Sigma$  in the following
form:
\begin{equation}\begin{split}
&\Sigma(y,x)=
-\Gamma^{(0)}_{\perp}(\underline{y},\underline{x}'\mid
\underline{z})\overline{R_{\perp}(\underline{z};V,J,M)G(\underline{x}',\underline{y}';V,J,M)|_{J=M=0}}G^{-1}(\underline{y}',\underline{x})\\&
-\Gamma^{(0)}_{\parallel}(\underline{y},\underline{x}'\mid
\underline{z}) K\left(%
\begin{array}{cc}
\underline{x}'& \underline{y}''  \\
  \underline{y}' & \underline{x}''\\
\end{array}%
\right)\Gamma^{(0)}_{\parallel}(\underline{y}'',\underline{x}''\mid
\underline{z}') D^{(0)}_{\parallel}(\underline{z},\underline{z}')
G^{-1}(\underline{y}',\underline{x})\\
&-\Gamma^{(0)}_{\perp}(\underline{y},\underline{x}'\mid
\underline{z}) K\left(%
\begin{array}{cc}
\underline{x}'& \underline{y}''  \\
  \underline{y}' & \underline{x}''\\
\end{array}%
\right)\Gamma^{(0)}_{\perp}(\underline{y}'',\underline{x}''\mid
\underline{z}') D^{(0)}_{\perp}(\underline{z},\underline{z}')
G^{-1}(\underline{y}',\underline{x})\\&-
\imath\Lambda K\left(%
\begin{array}{cc}
\textbf{r},\underline{t},j & \textbf{r},\underline{t}^I,j^I   \\
  \textbf{r}^{II},\underline{t}^{II},j^{II}  & \textbf{r},\underline{t}^I,j^I  \\
\end{array}%
\right)G^{-1}_{j^{II}
j'}(\textbf{r}^{II},\underline{t}^{II},\textbf{r}',\underline{t}').
\label{AvMOp2}
\end{split}\end{equation}
The first term in (\ref{AvMOp2}) is the Hartree term in the presence
of a disorder. The second and the third terms link the mass operator
to the average two-particle Green's function, and therefore, they
are the Fock contributions to the mass operator. By introducing the
vertex function $\Gamma_{\parallel(\perp)}$ :
\begin{equation}K^{(0)}\left(%
\begin{array}{cc}
  \underline{x} & \underline{y}'  \\
  \underline{y} & \underline{x}' \\
\end{array}%
\right)\Gamma_{\parallel(\perp)}(\underline{y}',\underline{x}'\mid \underline{z}')D_{\parallel(\perp)}(\underline{z}',\underline{z})= K\left(%
\begin{array}{cc}
  \underline{x} & \underline{y}'  \\
  \underline{y} & \underline{x}' \\
\end{array}%
\right)\Gamma^{(0)}_{\parallel(\perp)}(\underline{y}',\underline{x}'\mid
\underline{z}')D^{(0)}_{\parallel(\perp)}(\underline{z}',\underline{z}),
\label{Gamma}
\end{equation}
where the free two-particle propagator $K^{(0)}$ is given by:
\begin{equation}
K^{(0)}\left(%
\begin{array}{cc}
 \underline{x} & \underline{y}'  \\
  \underline{y} & \underline{x}' \\
\end{array}%
\right)=-\imath
G(\underline{x},\underline{y}')G(\underline{x}',\underline{y}),\label{TGF0V}
\end{equation}
one can rewrite the mass operator  in  the following form:
\begin{equation}\begin{split}&\Sigma(\underline{y},\underline{x})=
-\Gamma^{(0)}_{\perp}(\underline{y},\underline{x}'\mid
\underline{z})\overline{R_{\perp}(\underline{z};V,J,M)G(\underline{x}',\underline{y}';V,J,M)|_{J=M=0}}
G^{-1}(\underline{y}',\underline{x})\\&+\imath
\Gamma^{(0)}_{\parallel}(\underline{y},\underline{x}'\mid
\underline{z}')G(\underline{x}',\underline{y}')\Gamma_{\parallel}(\underline{y}',\underline{x}\mid
\underline{z}')D_{\parallel}(\underline{z},\underline{z}')+ \imath
\Gamma^{(0)}_{\perp}(\underline{y},\underline{x}'\mid
\underline{z}')G(\underline{x}',\underline{y}')\Gamma_{\perp}(\underline{y}',\underline{x}\mid
\underline{z}')D_{\perp}(\underline{z},\underline{z}')\\&-
\imath\Lambda K\left(%
\begin{array}{cc}
\textbf{r},\underline{t},j & \textbf{r},\underline{t}^I,j^I   \\
  \textbf{r}^{II},\underline{t}^{II},j^{II}  & \textbf{r},\underline{t}^I,j^I  \\
\end{array}%
\right)G^{-1}_{j^{II}
j'}(\textbf{r}^{II},\underline{t}^{II},\textbf{r}',\underline{t}').\label{Zitt1}\end{split}\end{equation}
\end{widetext}
By comparing the expression for the mass operator (\ref{MO1}) in the
absence of a symmetry-breaking disorder
  with Eq. (\ref{AvMOp2}), we find two differences. The
 first one is the presence of an additional term, $\Sigma_\Lambda =-\imath\Lambda KG^{-1}$. The second difference
 is related to the corresponding Hartree terms. In the presence of a symmetry-breaking disorder the
 Hartree term depends on the average of the
 product $\overline{R_{\perp}(\underline{z};V,J,M)G(\underline{x}',\underline{y}';V,J,M)}$ of two random
 functionals, defined by Eq. (\ref{2F}). \\
 Let us for a moment replace the average $\overline{R_{\perp}G}$ with the product of the averages $R_{\perp}.G$:
   \begin{equation}\overline{R_{\perp}(\underline{z};V,J,M)G(\underline{x},\underline{y};V,J,M)}|_{J=M=0}\rightarrow
   R_{\perp}(\underline{z})G(\underline{x},\underline{y}).\label{MapZ}\end{equation}
 Although the average of the product of two random quantities is not equal to the product of their averages,
 the replacement (\ref{MapZ}) greatly simplifies the equations and allows us to map them to the Zittartz's equations.
 Note, that in his work Zittartz took into account only the
lowest-order contribution from the
  disorder to the mass operator $\Sigma_\Lambda$, which corresponds to the replacement of  $K$
  by the free two-particle propagator $K^{(0)}$. In this approximation  we calculate for the mass operator:
  \begin{equation}\begin{split}
&\Sigma(\underline{y},\underline{x})=-\Gamma^{(0)}_\perp(\underline{y},\underline{x}|\underline{z})R_\perp(\underline{z})+
\delta(\textbf{r}-\textbf{r}')\Lambda
G_{j'j}(\textbf{r},\underline{t}',\textbf{r},\underline{t})\\
&+\imath \Gamma^{(0)}_{\parallel}(\underline{y},\underline{x}'\mid
\underline{z}')G(\underline{x}',\underline{y}')\Gamma_{\parallel}(\underline{y}',\underline{x}\mid
\underline{z}')D_{\parallel}(\underline{z},\underline{z}').\label{Zitt2}\end{split}\end{equation}
 The first and the third terms in (\ref{Zitt2}) represent the photonic and excitonic contributions to the order
 parameter. By nullifying the sources in Eq. (\ref{SDE1V})
 we obtain a relationship between the average photonic field $R_\perp$
and the average single-particle Green's function:
\begin{equation}
R_{\perp}(\underline{z})=\imath
D^{(0)}_{\perp}(\underline{z},\underline{z}')\Gamma^{(0)}_{\perp}(\underline{y},\underline{x}\mid
\underline{z}')G(\underline{x},\underline{y}). \label{SDE1Va}
\end{equation}
The last equation leads to an equation, similar to  Eq.
(\ref{Ord1}):
\begin{equation}R_\perp(\textbf{k})=\sum_{\textbf{q}}g(\textbf{k}-\textbf{q})\int
\frac{d \omega}{2\pi} G_{cv}(\textbf{q};\omega)
/\left(\Omega^2(\textbf{k})-\mu^2\right).
\label{Ord1a}\end{equation}
 The sum of photonic and excitonic contributions to the order parameter is:
\begin{equation}
\Delta(\textbf{k})=\sum_{\textbf{q}}\left[\frac{g(\textbf{k}-\textbf{q})}{\Omega^2(\textbf{k})-\mu^2}+
V(\textbf{k}-\textbf{q})\right]\int
\frac{d\omega}{2\pi}G_{cv}(\textbf{q},\omega).
  \label{Z1}\end{equation}
  Since our random potential has a variance (\ref{RP1}), the equation (21) of Zittartz assumes the following form:
\begin{equation}
\widetilde{G}_{ij}(\omega)=\Lambda\sum_{\textbf{q}}G_{ij}(\textbf{k},\omega).\label{Z2}
\end{equation}
The Fourier transform of the single-particle Green's function has
the form (Eq. (24) of Zittartz):
\begin{equation}\begin{split}
&\widehat{G}(\textbf{k},\omega)=\left(%
\begin{array}{cc}
 G_{cc}(\textbf{k},\omega) & G_{cv}(\textbf{k},\omega)  \\
  G_{cv}(\textbf{k},\omega) & G_{vv}(\textbf{k},\omega) \\
\end{array}%
\right)=-\frac{1}{D} \times\\&\left(%
\begin{array}{cc}
  \omega-\widetilde{G}_{vv}(\omega)+\eta(\textbf{k})+\imath 0^+ & \Delta(\textbf{k})+\widetilde{G}_{cv}(\omega)\\
  \Delta(\textbf{k})+\widetilde{G}_{cv}(\omega) & \omega-\widetilde{G}_{cc}(\omega)-\eta(\textbf{k})-\imath 0^+\\
\end{array}%
\right),\label{Z3}
\end{split} \end{equation}
where
\begin{equation}\begin{split}
&D=[\Delta(\textbf{k})+\widetilde{G}_{cv}(\omega)]^2-(\omega-\widetilde{G}_{cc}(\omega)-\eta(\textbf{k})+\imath
0^+)\times\\&(\omega-\widetilde{G}_{vv}(\omega)+\eta(\textbf{k})-\imath
0^+).\label{Z4}\end{split} \end{equation}
  Here, $\eta(\textbf{k})$ depends on the chemical potential, and is defined self-consistently  by
the solution of the following equation:
\begin{equation}\begin{split}
&\eta(\textbf{k})=\frac{1}{2}\left[\textit{E}_g+\frac{\pi^2}{2m_{exc}L^2}
+\frac{\textbf{k}^2}{2m_{exc}}-\mu\right]\\&-\frac{1}{2}
\sum_{\textbf{q}}V(\textbf{k}-\textbf{q})\int
\frac{d\omega}{2\pi}\left[G_{cc}(\textbf{q},\omega)+G_{vv}(\textbf{q},\omega)\right].\label{Z4}\end{split}\end{equation}
 Equations (\ref{Z1})-(\ref{Z4}) form a closed set of equations, that can be solved at any
 density. In the high density regime we can neglect the small corrections to $\eta$ due to $G_{cc}$ and
 $G_{vv}$, and map our equations (\ref{Z1})-(\ref{Z4}) to
 the zero-temperature version of Eq. (\ref{Zeqs}). As a result one
 could
 come up with the conclusion that the order parameter and the energy gap are gradually suppressed up to a critical disorder
 strength.\\
 Strictly speaking, $\overline{RG}\neq RG $, and therefore, all results obtained by using approximation
(\ref{MapZ}) should be considered questionable. More importantly,
the $\overline{RG}$ term does not allow us to prove the existence
of the Goldstone mode below the critical temperature, as we did in
Sec II. Going beyond the assumption (\ref{MapZ}) is a very
challenging task, which requires to take into account
diagrammatically irreducible vertex parts and an infinite number
of
 diagrams neglected by the assumption $\overline{RG}= RG $.  \\

\section{SUMMARY}
We have applied the CPT Green's function formalism to the problem of
cavity polaritons in the presence of a symmetry-breaking
   disorder. In contrast with the nonlinear sigma-model and the saddle-point equations, the Hartree term in the
    mass operator does not allow us to map
    the SD equations to the corresponding equations in the work by
    Zittartz.\\
    The saddle-point approximation and the
    replica trick could be responsible for the different expressions for the mass operator in the presence of a disorder. In the absence of
    a symmetry-breaking disorder, the saddle-point approximation
    leads not only to the correct  gap equation  in the high density
    regime, but by investigating the Gaussian fluctuations about the saddle point one can
    obtain the collective mode spectrum\cite{Melo,MSL} as well. Thus, we might
    suggest that the replica trick is not the right tool to perform the averaging over the random potential in the case
    of cavity polaritons in the presence of a symmetry-breaking
   disorder.

\end{document}